\documentclass[doublecol]{epl2} 

\newcommand{\bfF}{\mathbf{F}}

\newcommand{\hatbfz}{\hat{\mathbf{z}}}
\newcommand{\hatbfx}{\hat{\mathbf{x}}}
\newcommand{\hatbfy}{\hat{\mathbf{y}}}

\newcommand{\calGu}{\mathcal{G}_{u}}
\newcommand{\calH}{\mathcal{H}}
\newcommand{\calG}{\mathcal{G}}
\newcommand{\calQ}{\mathcal{Q}}
\newcommand{\calF}{\mathcal{F}}

\title{High-fidelity universal quantum gates through quantum interference}
\shorttitle{High-fidelity universal quantum gates through quantum interference} 

\author{Ran Li\inst{1,2,3} \and Frank Gaitan\inst{1,2,4}}
\shortauthor{Li and Gaitan}

\institute{                    
  \inst{1} Advanced Sciences Institute, The Institute of Physical and Chemical
Research (RIKEN), Wako-shi, Saitama, 351-0198, Japan\\
  \inst{2} CREST, Japan Science and Technology Agency (JST), Kawaguchi, 
Saitama, 332-0012, Japan\\
  \inst{3} Department of Physics, Kent State University, Stark Campus, North
Canton, OH 44720\\
  \inst{4} Department of Physics, Southern Illinois University, Carbondale,
IL 62901-4401
}

\abstract{
Twisted rapid passage is a type of non-adiabatic rapid passage that gives rise
to controllable quantum interference effects that were first observed 
experimentally in 2003. We show that twisted rapid passage sweeps can be used 
to implement a universal set of quantum gates that operate with high-fidelity. 
For each gate in the universal set, sweep parameter values are provided which 
simulations indicate will yield a quantum gate with error probability 
$P_{e}<10^{-4}$. Note that all gates in this universal set are driven by a 
\textit{single\/} type of control field (twisted rapid passage), and the error 
probability for each gate falls below the rough-and-ready estimate for 
the accuracy threshold $P_{a}\sim 10^{-4}$. The simulations suggest that the 
universal gate set produced by twisted rapid passage shows promise for use 
in a fault-tolerant scheme for quantum computing.}

\begin{document}

\maketitle

\section{Introduction}
\label{sec1}

The physical context for our discussion is the accuracy threshold theorem
\cite{att1,att2,att3,att4,att5,att6,att7,att8} which established that a quantum
computation of arbitrary duration could be done, with arbitrarily small error
probability, in the presence of noise, and using imperfect quantum gates,
under the following conditions. (1)~The computational data is protected by a
sufficiently layered concatenated quantum error correcting code. 
(2)~Fault-tolerant protocols for quantum computation, error correction, and
measurement are used. (3)~A universal set of unencoded quantum gates is 
available with the property that each gate in the set has an error probability
$P_{e}$ that falls below a value $P_{a}$ known as the accuracy threshold. The
value of the threshold is model-dependent, though for many, $P_{a}\sim
10^{-4}$ has become a rough-and-ready estimate. Thus gates are anticipated to
be approaching the accuracies needed for fault-tolerant quantum computing
when $P_{e}<10^{-4}$. One of the principal challenges facing the field of
quantum computing is finding a way to implement a universal set of unencoded
quantum gates for which all gate error probabilities satisfy $P_{e}< 10^{-4}$.

In this Letter numerical simulation results are presented which suggest that a
class of non-adiabatic rapid passage sweeps, first realized experimentally in
1991 \cite{zwan1}, and known as twisted rapid passage (TRP), should be capable
of implementing a universal set of unencoded quantum gates $\mathcal{G}_{u}$
that operate non-adiabatically, and with gate error probabilities satisfying
$P_{e}<10^{-4}$. $\mathcal{G}_{u}$ consists of the one-qubit Hadamard and
NOT gates, together with variants of the one-qubit $\pi /8$ and phase gates,
and the two-qubit controlled-phase gate. The universality of $\mathcal{G}_{u}$
was demonstrated in Ref.~\cite{lhg2}. This level of gate accuracy is due to
controllable quantum interference effects that arise during a TRP sweep
\cite{fg1,lhg1}, and which were observed using NMR in 2003 \cite{zwan2}. 
To find sweep parameter values that yield such high-performance gates, it
proved necessary: (i)~to combine the simulations with an optimization procedure
that searches for minima of $P_{e}$ \cite{lhg2,lhg1}; and (ii)~for the
modified controlled-phase gate, to also apply the symmetrized evolution of
Ref.~\cite{zan}. 

The outline of this Letter is as follows. We begin with a summary of the
essential properties of TRP. This is followed by a discussion of how 
the simulation and optimization are done, and how symmetrized evolution is 
incorporated into the two-qubit dynamics. We then present our simulation
results for the different gates in $\mathcal{G}_{u}$. We close with a
discussion of our results and of future work.

\section{Preliminaries}
\label{sec2}

To introduce TRP \cite{fg1,lhg1}, we consider a single qubit interacting with
an external control field $\mathbf{F}(t)$ via the Zeeman interaction 
$H_{z}(t) = -\mbox{\boldmath{$\sigma$}}\cdot \mathbf{F}(t)$, where the
$\sigma_{i}$ are the Pauli matrices ($i=x,y,z$). TRP is a generalization of
adiabatic rapid passage (ARP). In ARP, the field $\bfF (t)$ is slowly inverted 
over a time $T_{0}$ such that $\bfF (t) = at\hatbfz + b\hatbfx$. In TRP, the 
control field is allowed to twist in the $x$-$y$ plane with time-varying
azimuthal angle $\phi (t)$, while simultaneously undergoing inversion along
the $z$-axis:
\begin{equation}
\bfF (t) = at\hatbfz + b\cos\phi (t)\hatbfx + b\sin\phi (t)\hatbfy .
\label{TRPcontrol}
\end{equation} 
Here $-T_{0}/2\leq t\leq T_{0}/2$, and the TRP inversion can be non-adiabatic.

\subsection{Controllable Quantum Interference}
\label{sec2a}

As shown in Ref.~\cite{lhg1}, the qubit undergoes resonance when $at -\hbar
\dot{\phi}(t)/2 = 0$. For polynomial twist with $\phi_{n}(t) = (2/n) Bt^{n}$,
this condition has $n-1$ roots, though only real-valued roots correspond to
resonance. Ref.~\cite{fg1} showed that for $n\geq 3$, the qubit undergoes
resonance multiple times during a \textit{single\/} TRP sweep: (i)~for all
$n\geq 3$, when $B>0$; and (ii) for odd $n\geq 3$, when $B<0$. For the 
remainder of this Letter we restrict ourselves to $B>0$, and to quartic twist
for which $n=4$ in $\phi_{n}(t)$. For quartic twist, the qubit passes
through resonance at the times $t=0,\pm\sqrt{a/\hbar B}$ \cite{fg1}. Thus the
time separating the qubit resonances can be controlled through variation of
the sweep parameters $B$ and $a$. Ref.~\cite{fg1} showed that these multiple
resonances have a strong influence on the qubit transition probability,
allowing transitions to be strongly enhanced or suppressed through a small
variation of the sweep parameters. Ref.~\cite{fg2} calculated the qubit
transition amplitude to all orders in the non-adiabatic coupling. The result
found there can be re-expressed as the following diagrammatic series:
\begin{equation}
\setlength{\unitlength}{0.05in}
T_{-}(t) = \begin{picture}(10,4)
              \put(10,-1.5){\vector(-1,0){3.25}}
              \put(5,-1.5){\line(1,0){1.75}}
              \put(5,-1.5){\vector(0,1){3.25}}
              \put(5,1.75){\line(0,1){1.75}}
              \put(5,3.5){\vector(-1,0){3.25}}
              \put(0,3.5){\line(1,0){1.75}}
           \end{picture}
\hspace{0.05in} +
\begin{picture}(20,4)
              \put(20,-1.5){\vector(-1,0){3.25}}
              \put(15,-1.5){\line(1,0){1.75}}
              \put(15,-1.5){\vector(0,1){3.25}}
              \put(15,1.75){\line(0,1){1.75}}
              \put(15,3.5){\vector(-1,0){3.25}}
              \put(10,3.5){\line(1,0){1.75}}
              \put(10,3.5){\vector(0,-1){3.25}}
              \put(10,-1.5){\line(0,1){1.75}}
              \put(10,-1.5){\vector(-1,0){3.25}}
              \put(5,-1.5){\line(1,0){1.75}}
              \put(5,-1.5){\vector(0,1){3.25}}
              \put(5,1.75){\line(0,1){1.75}}
              \put(5,3.5){\vector(-1,0){3.25}}
              \put(0,3.5){\line(1,0){1.75}}
           \end{picture}
\hspace{0.05in} + \hspace{0.05in} \cdots \hspace{0.05in} .
\label{diagser}
\end{equation} 
Lower (upper) lines correspond to propagation in the negative (positive)
energy level, and the vertical lines correspond to transitions between the
two energy levels. The calculation sums the probability amplitudes for all
interfering alternatives that allow the qubit to end up in the positive
energy level at time $t$ given that it was initially in the negative energy
level. As we have seen, varying the TRP sweep parameters varies the time
separting the resonances. This in turn changes the value of each diagram in
eq.~(\ref{diagser}), and thus alters the interference between alternatives in
the quantum superposition, It is the sensitivity of the individual 
alternatives/diagrams to the time separation of the resonances that allows TRP 
to manipulate this quantum interference. Zwanziger et al.\ \cite{zwan2} 
observed these interference effects in the transition probability using NMR 
and found quantitative agreement between theory and experiment. It is the link
between the TRP sweep parameters and this quantum interference that we believe
makes it possible for TRP to drive highly accurate non-adiabatic one- and
two-qubit gates.

\subsection{Simulation and Optimization}
\label{sec2b}

A detailed presentation of our simulation and optimization protocols appears 
in Refs.~\cite{lhg2,lhg1}. We can only give a brief sketch of that presentation
here. As is well-known, the Schrodinger dynamics applies a unitary 
transformation $U$ to an initial quantum state $|\psi\rangle$ which is driven 
by the system Hamiltonian $H(t)$. The Hamiltonian (see below) is assumed to 
contain terms
that Zeeman-couple each qubit to the TRP control field $\bfF (t)$. Assigning
values to the TRP sweep parameters ($a$,$b$,$B$,$T_{0}$) determines $H(t)$,
which then determines the actual unitary transformation $U_{a}$ applied. The
task is to find sweep parameter values that produce a $U_{a}$ that approximates
a target gate $U_{t}$ sufficiently closely that its error probability (defined
below) satisfies $P_{e}<10^{-4}$. In the following, the target gate $U_{t}$
will be one of the gates in the universal set $\mathcal{G}_{u}$. Since
$\calGu$ contains only one- and two-qubit gates, our simulations will only
involve one- and two-qubit systems. For the one-qubit simulations, the
Hamiltonian $H_{1}(t)$ is the Zeeman Hamiltonian introduced earlier.
Ref.~\cite{lhg2} showed that it can be written in the following dimensionless 
form: 
\begin{equation}
\calH_{1}(\tau ) = (1/\lambda )\left\{ -\tau\sigma_{z} -\cos\phi_{4}(\tau )
                     \sigma_{x} -\sin\phi_{4}(\tau )\sigma_{y} \right\} . 
\label{oneqbtHam}
\end{equation}
Here: $\tau = (a/b)t$; $\lambda = \hbar a/b^{2}$; and for quartic twist,
$\phi_{4}(\tau ) = (\eta_{4}/2\lambda )\tau^{4}$ with $\eta_{4} =\hbar Bb^{2}/
a^{3}$. For the two-qubit simulations, the Hamiltonian $H_{2}(t)$ contains
terms that Zeeman-couple each qubit to the TRP control field, and an Ising
interaction that couples the two qubits. Note that alternative two-qubit
interactions can easily be considered, though we focus on the Ising interaction
here. The energy-levels for the resulting Hamiltonian contain a 
resonance-frequency degeneracy that was found to spoil gate performance.
Specifically, the resonance-frequency for transitions between the ground-
and first-excited states ($E_{1}\leftrightarrow E_{2}$) is the same as that
for transitions between the second- and third-excited states ($E_{3}
\leftrightarrow E_{4}$). To remove this degeneracy a term $c_{4}|E_{4}(\tau )
\rangle\langle E_{4}(\tau )|$ was added to $H_{2}(t)$. Combining all these
remarks, one arrives at the following dimensionless two-qubit Hamiltonian
\cite{lhg2}:
\begin{eqnarray}
\lefteqn{\calH_{2}(\tau )  = } \nonumber\\
 & & [-(d_{1}+d_{2})/2 +\tau /\lambda ]\sigma^{1}_{z}
                        -(d_{3}/\lambda )[\cos\phi_{4}\sigma^{1}_{x} +
                         \sin\phi_{4}\sigma_{y}^{1}] \nonumber\\
  &  & \hspace{0.2in}+[-d_{2}/2 +\tau /\lambda ]\sigma_{z}^{2} -(1/\lambda )
         [\cos\phi_{4}
         \sigma_{x}^{2}+\sin\phi_{4}\sigma_{y}^{2} ] \nonumber\\
  &  & \hspace{0.40in} -(\pi d_{4}/2)\sigma_{z}^{1}\sigma_{z}^{2} + c_{4}
        |E_{4}(\tau )\rangle
         \langle E_{4}(\tau )| .
\label{twoqbtHam}
\end{eqnarray}
Here: (1)~$b_{i} = \hbar\gamma_{i}B_{rf}/2$, $\omega_{i}=\gamma_{i}B_{0}$, and
$i=1,2$; (2)~$\tau = (a/b_{2})t$, $\lambda = \hbar a/b_{2}^{2}$, and $\eta_{4}
=\hbar Bb_{2}^{2}/a^{3}$; and (3)~$d_{1}=(\omega_{1}-\omega_{2})b_{2}/a$, 
$d_{2}=(\Delta /a)b_{2}$, $d_{3}=b_{1}/b_{2}$, and $d_{4}=(J/a)b_{2}$, where
$\Delta$ is a detuning parameter \cite{lhg2}.

The numerical simulation assigns values to the TRP sweep parameters and then
integrates the Schrodinger equation to obtain the unitary transformation
$U_{a}$ produced by the sweep. To assess how closely $U_{a}$ approximates the
target gate $U_{t}$, it proves useful to introduce the positive operator $P=
(U_{a}^{\dagger}-U_{t}^{\dagger})(U_{a}-U_{t})$. Given $U_{a}$, $U_{t}$, and
an initial state $|\psi\rangle$, one can work out the error probability $P_{e}
(\psi )$ for the TRP final state $|\psi_{a}\rangle = U_{a}|\psi\rangle$,
relative to the target final state $|\psi_{t}\rangle = U_{t}|\psi\rangle$. The
gate error probability $P_{e}$ is defined to be the worst-case value of
$P_{e}(\psi )$: $P_{e}\equiv \max_{|\psi\rangle}P_{e}(\psi )$. 
Ref.~\cite{lhg1} showed that $P_{e}$ satisfies the bound $P_{e}\leq
Tr\, P$, where the RHS is the trace of the positivie operator $P$ introduced
above. Once $U_{a}$ is known, $Tr\, P$ is easily evaluated, and so it makes
a convenient proxy for $P_{e}$, which is harder to calculate. To find TRP
sweep parameter values that yield highly accurate non-adiabatic quantum gates,
it proved necessary to combine the numerical simulations with function
minimization algorithms that search for sweep parameter values that minimize
the $Tr\, P$ upper bound \cite{numrec}. The multi-dimensional downhill simplex
method was used for the one-qubit gates, while simulated annealing was used
for the two-qubit modified controlled-phase gate. This produced the one-qubit
gate results that will be presented below. However, for the modified
controlled-phase gate, simulated annealing was only able to find parameter
values that gave $P_{e}\leq 1.27\times 10^{-3}$ \cite{lhg2}. To further 
improve the performance of this two-qubit gate, it proved necessary to 
incorporate the symmetrized evolution of Ref.~\cite{zan} to obtain a modified
controlled-phase gate with $P_{e}<10^{-4}$. We now briefly describe how
symmetrized evolution is incorporated into our simulations. 

\subsection{Symmetrized Evolution and TRP}
\label{sec2c}

Ref.~\cite{zan} introduced a unitary group-symmetrization procedure that
yields an effective dynamics that is invariant under the action of a finite
group $\calG$. We incorporate this group-symmetrization into a TRP sweep by
identifying the group $\calG$ with a finite symmetry group of the target gate
$U_{t}$, and then applying the procedure of Ref.~\cite{zan} to filter out the
$\calG$-noninvariant part of the TRP dynamics. As the $\calG$-noninvariant 
dynamics is manifestly bad dynamics relative to $U_{t}$, group-symmetrized
TRP yields a better approximation to $U_{t}$. We briefly describe the 
group-symmetrization procedure, and then show how it can be incorporated into
a TRP sweep.

Consider a quantum system $\calQ$ with time-independent Hamiltonian $H$ and
Hilbert space $\calH$. The problem is to provide $\calQ$ with an effective
dynamics that is invariant under a finite group $\calG$, even when $H$ itself
is not $\calG$-invariant. This symmetrized dynamics manifests as a
$\calG$-invariant effective propagator $\tilde{U}$ that evolves the system
state over a time $t$. Let $\{ \rho_{i}=\rho (g_{i})\}$ be a unitary
representation of $\calG$ on $\calH$, and let $|\calG |$ denote the order of
$\calG$. The procedure begins by partitioning the time-interval ($0$,$t$) into
$N$ subintervals of duration $\Delta t_{N}=t/N$, and then further partitioning
each subinterval into $|\calG |$ smaller intervals of duration $\delta t_{N}
=\Delta t_{N}/|\calG |$. Let $\delta U_{N} =\exp\left[ -(i/\hbar )\delta t_{N}
H\right]$ denote the $H$-generated propagator for a time-interval $\delta 
t_{N}$, and assume that the time to apply each $\rho_{i}\in\calG$ is negligible
compared to $\delta t_{N}$ (bang-bang limit \cite{viollyd}). In each 
subinterval, the following sequence of transformations is applied: $U(\Delta
t_{N}) = \prod_{i=1}^{|\calG |}\rho_{i}^{\dagger}\delta U_{N}\rho_{i}$.
Ref.~\cite{zan} showed that: (i)~$U(\Delta t_{N})\rightarrow \exp [-(i/\hbar )
\Delta t_{N}\tilde{H}]$ as $N\rightarrow \infty$, where $\tilde{H} = 
(1/|\calG |)\sum_{i=1}^{|\calG |}\rho_{i}^{\dagger}H\rho_{i}$; (ii)~$\tilde{H}$
is $\calG$-invariant ($[\tilde{H},\rho_{i}]=0$ for all $\rho_{i}\in\calG$);
and (iii)~the propagator $\tilde{U}$ over ($0$,$t$) is $\tilde{U}=\exp [-(i/
\hbar )t\tilde{H}]$, which is $\calG$-invariant due to the $\calG$-invariance
of $\tilde{H}$. The end result is an effective propagator $\tilde{U}$ that is
$\calG$-invariant as desired.

This procedure can be generalized to allow for a time-dependent Hamiltonian
$H(t)$. To do this, the time interval ($0$,$t$) must be divided into
sufficiently small subintervals that $H(t)$ is effectively constant in each.
Within each subinterval, the above time-independent argument is applied, 
yielding a $\calG$-symmetrized propagator for that subinterval. Combining the
effective propagators for each of the subintervals then gives the full
propagator $\tilde{U} = T[\exp (-i/\hbar\int_{0}^{t}d\tau \tilde{H}(\tau ))]$,
where $T$ indicates a time-ordered exponential, and $\tilde{H}(t)= (1/|\calG |)
\sum_{i=1}^{|\calG |}\rho_{i}^{\dagger}H(t)\rho_{i}$.

For our two-qubit simulations, the target gate is the modified controlled-phase
gate $V_{cp}=(1/2)[(I^{1} +\sigma^{1}_{z})I^{2} -(I^{1}-\sigma_{z}^{1})
\sigma_{z}^{2}]$ which is invariant under the group $\calG = \{ I^{1}I^{2},
\sigma_{z}^{1},\sigma_{z}^{2},\sigma_{z}^{1}\sigma_{z}^{2}\}$. Thus $|\calG |
=4$, and we set $\rho_{1}=I^{1}I^{2}, \ldots , \rho_{4}=\sigma_{z}^{1}
\sigma_{z}^{2}$. Switching over to dimensionless time, we partition the sweep
time-interval ($-\tau_{0}/2,\tau_{0}/2$) into sufficiently small subintervals
that our two-qubit Hamiltonian $H_{2}(\tau )$ is effectively constant within
each. We then apply the time-independent symmetrization procedure to each
subinterval with the $V_{cp}$ symmetry group acting as $\calG$. Combining the
effective propagators for each of the subintervals as above gives the
$\calG$-symmetrized propagator for the full TRP sweep $\tilde{U}= T[\exp
(-i/\hbar )\int_{-\tau_{0}/2}^{\tau_{0}/2} d\tau \tilde{H}(\tau ))]$, with
$\tilde{H}(\tau ) = (1/4)\sum_{i=1}^{4}\rho_{i}^{\dagger}H_{2}(\tau )\rho_{i}$.
We shall see that $\calG$-symmetrized TRP yields an approximation to $V_{cp}$
with $P_{e}<10^{-4}$.

\section{Simulation Results}
\label{sec3}

\subsection{One-qubit gates}
\label{sec3a}

Operator expressions for the target gates are: 
(i)~\textit{Hadamard\/}---$U_{h}=(1/\sqrt{2})(\sigma_{z}+\sigma_{x})$;
(ii)~\textit{NOT\/}---$U_{not}=\sigma_{x}$;
(iii)~\textit{modified $\pi /8$\/}---$V_{\pi /8}=\cos (\pi /8)\sigma_{x}-
\sin (\pi /8)\sigma_{y}$; and 
(iv)~\textit{modified Phase\/}---$V_{p}=(1/\sqrt{2})(\sigma_{x}-\sigma_{y})$.
The gate fidelity is calculated using $\calF_{n} = (1/2^{n}) Re[Tr(
U^{\dagger}_{a}U_{t})]$, where $n$ is the number of qubits acted on by the 
gate. This fidelity is especially convenient as it is related to our $Tr\, P$
upper bound: $\calF_{n} = 1 -(1/2^{n+1})Tr\, P$ \cite{lhg2}. Finally, the
connection between the TRP experimental and theoretical parameters is given
in Refs.~\cite{lhg2}, \cite{fg1}, and \cite{lhg1} for superconducting, NMR, 
and atom-based qubits, respectively.

A study of the TRP-implementation of these one-qubit gates was first reported
in Ref.~\cite{lhg1}. It proves useful to reparameterize the TRP sweep 
parameters $\lambda\rightarrow\lambda^{\ast}$ and $\eta_{4}\rightarrow
\eta_{4}^{\ast}$, where $\lambda =\lambda^{\ast}\exp [ -(\lambda^{\ast}-
\lambda^{0})/\lambda^{0}]$ and $\eta_{4}=\eta_{4}^{\ast}\exp [-(\eta_{4}^{\ast}
-\eta_{4}^{0})/\eta_{4}^{0}]$. For each one-qubit gate in $\calGu$, the
fixed-point of the reparameterization ($\lambda^{0}$,$\,\eta_{4}^{0}$) is
given by the optimum sweep parameter values found in Ref.~\cite{lhg1}. 
Table~\ref{table1} presents the values for the optimum sweep parameters 
($\lambda^{\ast}$,$\,\eta_{4}^{\ast}$) that produced our best results for 
$Tr\, P$ for each of the gates in $\calGu$.\hspace{-0.1in} 
\begin{largetable}
\caption{Simulation results for the one-qubit gates in $\calGu$. The gate error
probability $P_{e}$ satisfies $P_{e}\leq Tr\, P$.}
\label{table1}
\begin{center}
\begin{tabular}{ccc|c|cc}\hline
Gate & $\lambda^{\ast}$ & $\eta_{4}^{\ast}$ & $Tr\, P$ & $\lambda^{0}$ &
 $\eta_{4}^{0}$ \\\hline
$U_{h}$ & $5.85$ & $2.93\times 10^{-4}$ & $9.30\times 10^{-6}$ & $5.8511$ &
 $2.9280\times 10^{-4}$\\
$U_{not}$ & $7.32$ & $2.93\times 10^{-4}$ & $1.12\times 10^{-5}$ & $7.3205$ &
 $2.9277\times 10^{-4}$\\
$U_{\pi /8}$ & $6.02$ & $8.15\times 10^{-4}$ & $3.55\times 10^{-5}$ & $6.0150$ 
&  $8.1464\times 10^{-4}$\\
$U_{p}$ & $5.98$ & $3.81\times 10^{-4}$ & $8.70\times 10^{-5}$ & $5.9750$ 
&  $3.8060\times 10^{-4}$\\
\end{tabular}
\end{center}
\end{largetable}
The fixed-point ($\lambda^{0}$,$\,\eta_{4}^{0}$) for each gate is also listed. 
In all one-qubit simulations,
the dimensionless inversion time was $\tau_{0}=80.0$. Since $P_{e}\leq Tr\, P$,
we see that $P_{e}<10^{-4}$ for all one-qubit gates in $\calGu$. The $Tr\, P$
values yield the following gate fidelities: (i)~Hadamard---$\calF_{h} =
0.9999\, 98$; (ii)~NOT---$\calF_{not}=0.9999\, 97$; (iii)~modified 
$\pi /8$---$\calF_{\pi /8} =0.9999\, 91$; and (iv)~modified Phase---$\calF_{p}
=0.9999\, 78$. In Table~\ref{table2}\hspace{-0.1in} 
\begin{largetable}
\caption{Variation of $Tr\, P$ for the Hadamard gate when the TRP sweep 
parameters are altered slightly from their optimum values. Variation of
$Tr\, P$ for the other one-qubit gates in $\calGu$ is comparable
to that of the Hadamard gate and so corresponding Tables for these other gates
are not shown.}
\label{table2}
\begin{center}
\begin{tabular}{ccc|ccc}\hline
$\lambda^{\ast}$ & $\eta_{4}^{\ast}$ & $Tr\, P$ & $\eta_{4}^{\ast}$ &
 $\lambda^{\ast}$ & $Tr\, P$ \\ \hline
$5.85$ & $2.92\times 10^{-4}$ & $2.24\times 10^{-5}$ & 
 $2.93\times 10^{-4}$ & $5.84$ & $1.24\times 10^{-5}$\\
       & $2.93\times 10^{-4}$ & $9.30\times 10^{-6}$ & 
                      & $5.85$ & $9.30\times 10^{-6}$\\
       & $2.94\times 10^{-4}$ & $6.06\times 10^{-5}$ & 
                      & $5.86$ & $1.12\times 10^{-5}$\\
\end{tabular}
\end{center}
\end{largetable}
we show how $Tr\, P$ varies when
$\lambda^{\ast}$ and $\eta_{4}^{\ast}$ are varied slightly from their optimum
values for the Hadamard gate. Similar behavior occurs for the other one-qubit 
gates in $\calGu$, and so in the interests of brevity, corresponding Tables 
for these other gates are not shown. We see that gate performance is a 
slowly-varying function of the sweep parameters 
($\lambda^{\ast}$,$\,\eta_{4}^{\ast}$).

\subsection{Modified controlled-phase gate}
\label{sec3b}

We complete the universal set $\calGu$ by presenting our simulation results for
the $\calG$-symmetrized TRP implementation of the modified controlled-phase
gate $V_{cp}$. In the two-qubit computational basis (eigenstates of 
$\sigma_{z}^{1}\sigma_{z}^{2}$), $V_{cp} = diag(1,1,-1,1)$. TRP implementation
of $V_{cp}$ \textit{without\/} symmetrized evolution was reported in
Ref.~\cite{lhg2}. The results presented there are superceded by the
$\calG$-symmetrized TRP results presented below. For purposes of later
discussion, note that the parameters appearing in $\calH_{2}(\tau )$ fall into
two sets. The first consists of the TRP sweep parameters 
($\lambda$,$\,\eta_{4}$,$\,\tau_{0}$), while the second set 
($c_{4}$,$\, d_{1}$,\ldots ,$\, d_{4}$) consists of parameters for 
degeneracy-breaking, detuning, and coupling. We partitioned the TRP sweep into
$N_{seq}=2500$ pulse sequences, with each sequence based on the four element
symmetry group for $V_{cp}$ introduced earlier. 

For the modified controlled-phase gate $V_{cp}$, gate performance was not
found to be very sensitive to small variations of the TRP sweep parameters. 
Instead, for $V_{cp}$ \textit{without\/} symmetrized evolution \cite{lhg2}, 
gate performance was most sensitive to $c_{4}$, $d_{1}$, and $d_{4}$. However, 
when symmetrized evolution was incorporated into the TRP sweep, $d_{1}$ ceased 
to be a critical parameter. Thus it only proved necessary to reparameterize 
$c_{4}\rightarrow c_{4}^{\ast}$ and $d_{4}\rightarrow d_{4}^{\ast}$, where
$c_{4}=c_{4}^{\ast}\exp [-(c_{4}^{\ast}-c_{4}^{0})/c_{4}^{0}]$ and
$d_{4} = d_{4}^{\ast}\exp [-(d_{4}^{\ast}-d_{4}^{0})/d^{0}_{4}]$. Simulations
incorporating symmetrized evolution determined the reparameterization 
fixed-point to be $c_{4}^{0}=2.173$, $d_{4}^{0}=0.8347$. The (optimized) 
parameter values $\lambda = 5.04$, $\eta_{4}=3.0\times 10^{-4}$, $\tau_{0}=
120$, $d_{1}=99.3$, $d_{2}=0.0$, $d_{3}=-0.41$, $d_{4}^{\ast}=0.835$, and
$c_{4}^{\ast} = 2.17$ produced a gate $U_{a}$ for which $Tr\, P = 8.87\times
10^{-5}$, gate fidelity $\calF_{cp}=0.9999\, 89$, and gate error probability
satisfying $P_{e}\leq 8.87\times 10^{-5}$. We see that by adding symmetrized
evolution to a TRP sweep we obtain an approximation to $V_{cp}$ with $P_{e}
\leq 10^{-4}$. In Table~\ref{table3}\hspace{-0.05in} 
\begin{largetable}
\caption{Variation of $Tr\, P$ for the modified controlled-phase gate when
$c_{4}^{\ast}$ and $d_{4}^{\ast}$ are altered slightly from their optimum
values.}
\label{table3}
\begin{center}
\begin{tabular}{ccc|ccc}\hline
$c_{4}^{\ast}$ & $d_{4}^{\ast}$ & $Tr\, P$ & $d_{4}^{\ast}$ & $c_{4}^{\ast}$
 & $Tr\, P$ \\\hline
$2.17$ & $0.834$ & $8.77\times 10^{-5}$ &
  $0.835$ & $2.16$ & $8.15\times 10^{-5}$ \\
          & $0.835$ & $8.77\times 10^{-5}$ &
          & $2.17$ & $8.77\times 10^{-5}$ \\
          & $0.836$ & $8.77\times 10^{-5}$ &
          & $2.18$ & $8.44\times 10^{-5}$ 
\end{tabular}
\end{center}
\end{largetable}
we show how $Tr\, P$ varies when the parameters $c_{4}^{\ast}$ and 
$d_{4}^{\ast}$ vary slightly from their optimum values. Sensitivity of gate
performance to the remaining parameters is comparable to that of 
$c_{4}^{\ast}$ and $d_{4}^{\ast}$ and so corresponding tables are not shown. 
We see that gate performance is a slowly-varying function of the parameters 
$c_{4}^{\ast}$ and $d_{4}^{\ast}$, as well as of 
($\lambda$,$\,\eta_{4}$,$\,\tau_{0}$) and ($d_{1}$,$\, d_{2}$,$\, d_{3}$).

\section{Discussion}
\label{sec4}

We have presented simulation results which suggest that TRP sweeps should be
capable of implementing a universal set of quantum gates $\calGu$ that
operate non-adiabatically and with gate error probability satisfying $P_{e}
<10^{-4}$. We note that all gates in the universal set $\calGu$ are driven
by a \textit{single\/} type of control field (TRP), and that the gate error
probability for all gates in $\calGu$ falls below the rough-and-ready estimate
for the accuracy threshold $P_{a}\sim 10^{-4}$. The simulation results
presented in this Letter suggest that the universal quantum gate set $\calGu$
produced by TRP shows promise for use in a fault-tolerant scheme for quantum
computing. Refs.~\cite{lhg2,fg1,lhg1} have shown how TRP sweeps can be applied 
to NMR, atomic, and superconducting qubits. It should also be possible to 
apply them to spin-based qubits in quantum dots using a magnetic field since 
the same Zeeman-coupling acts as with NMR qubits. Although we have studied a 
number of forms of polynomial twist, as well as periodic twist \cite{lg1}, we 
have found that quartic twist provides best all-around performance when it 
comes to making the gates in $\calGu$. At present we do not have arguments 
that explain why quartic twist works better than other forms of twist. 
We are currently working to develop a theory of the optimum twist profile in 
an effort to understand this question.

\acknowledgments
We thank: (i)~F. Nori, RIKEN, and CREST for making our visit to RIKEN possible;
(ii)~RIKEN for access to the RIKEN Super Combined Cluster on which the 
simulations incorporating symmetrized evolution were done; and (iii)~T. Howell
III for continued support.

\end{document}